\documentclass[letterpaper, 10 pt, conference]{ieeeconf}
\usepackage{cite}
\usepackage{amsmath,amssymb,amsfonts}
\usepackage{algorithmic}
\usepackage{graphicx}
\usepackage{textcomp}
\usepackage{wrapfig}
\usepackage{xcolor}

\IEEEoverridecommandlockouts                              

\title{\LARGE \bf
Activity-Aware Deep Cognitive Fatigue Assessment using Wearables
}

\author{Mohammad Arif Ul Alam
\thanks{Department of Computer Science, University of Massachusetts Lowell
        {\tt\small mohammadariful\_alam@uml.edu}}%
}

\begin{document}

\maketitle
\thispagestyle{empty}
\pagestyle{empty}

\begin{abstract}
Cognitive fatigue has been a common problem among workers which has become an increasing global problem since the emergence of COVID-19 as a global pandemic. While existing multi-modal wearable sensors-aided automatic cognitive fatigue monitoring tools have focused on physical and physiological sensors (ECG, PPG, Actigraphy) analytic on specific group of people (say gamers, athletes, construction workers), activity-awareness is utmost importance due to its different responses on physiology in different person. In this paper, we propose a novel framework, Activity-Aware Recurrent Neural Network (\emph{AcRoNN}), that can generalize individual activity recognition and improve cognitive fatigue estimation significantly. We evaluate and compare our proposed method with state-of-art methods using one real-time collected dataset from 5 individuals and another publicly available dataset from 27 individuals achieving max. 19\% improvement.
\end{abstract}


\begin{figure*}[ht!]
\begin{minipage}{0.20\textwidth}
 \begin{center}
   \includegraphics[width=\linewidth]{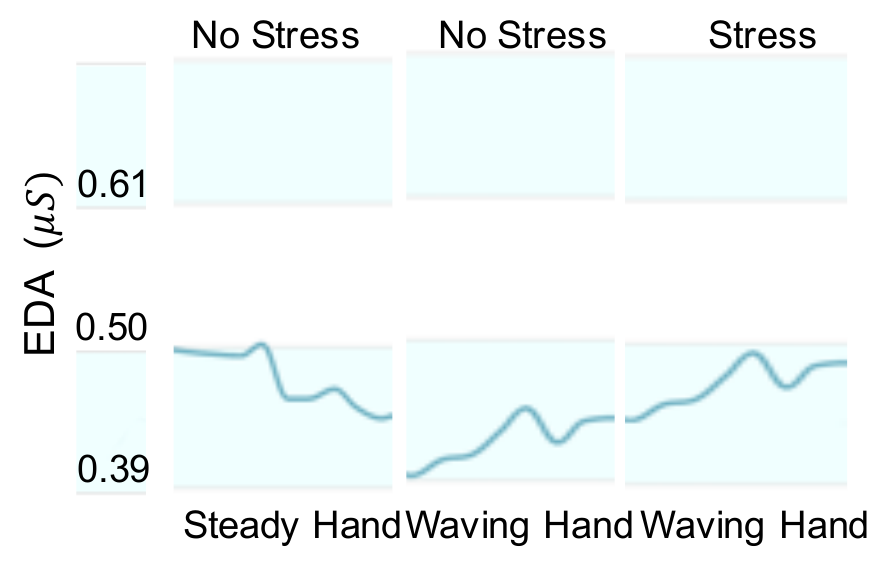}
   \caption{Electrodermal activity (EDA) responses (micro Siemens $\mu S$) on 2 seconds of 2 dominant hand gestures: steady hand and waving hand twice during two physiological states: stress and no stress of the same subject. Note that, the subject was in sitting position.}
   \label{fig:example}
 \end{center}
\end{minipage}
\begin{minipage}{0.05\textwidth}
\end{minipage}
\begin{minipage}{0.75\textwidth}
 \begin{center}
   \includegraphics[width=\linewidth]{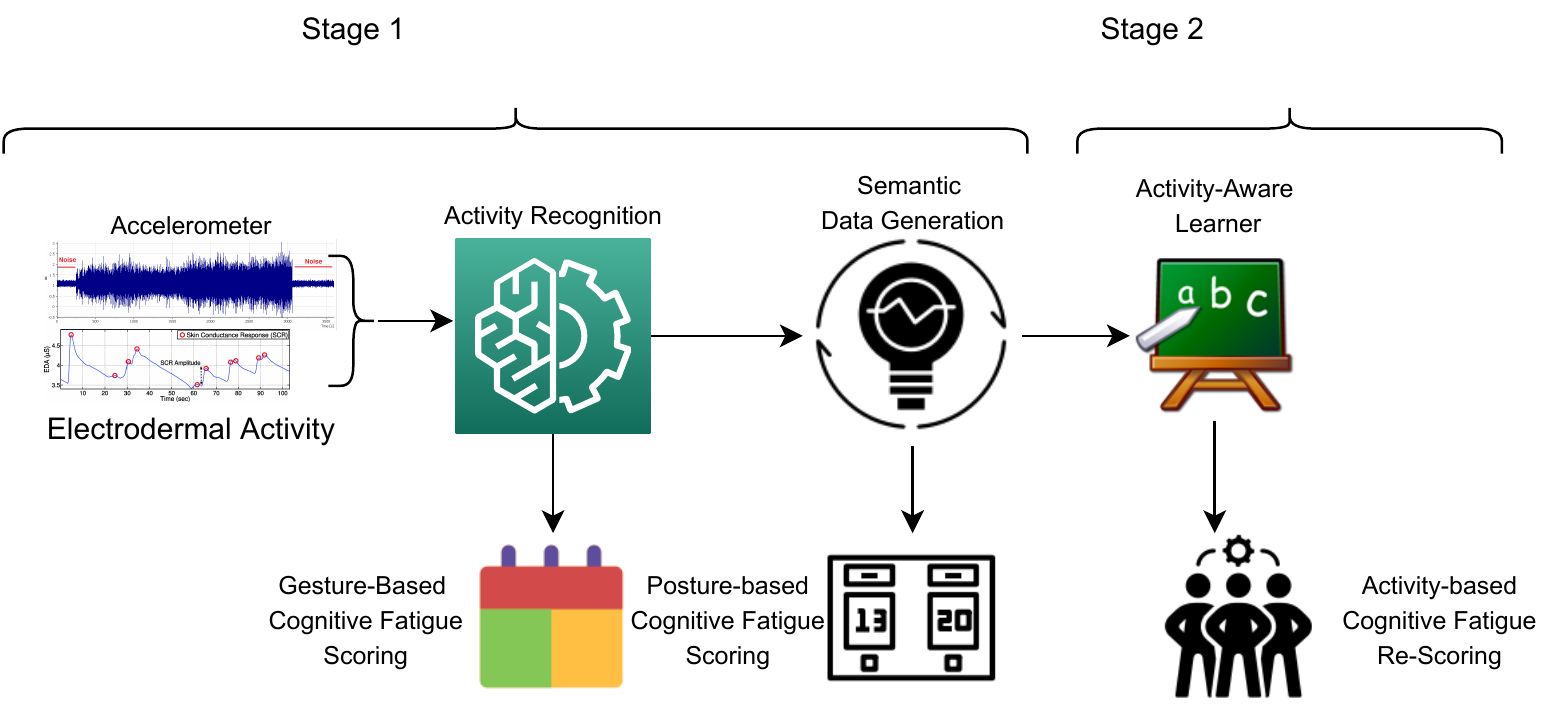}
   \caption{High-level schematic diagram of \emph{AcRoNN} architecture that consists two stages of learning}
   \label{fig:AcRoNN}
 \end{center}
\end{minipage}
\end{figure*}

\section{Introduction}
Cognitive fatigue is a syndrome conceptualized as resulting from chronic workplace stress that has not been successfully managed \cite{who19}. Although, cognitive fatigue is not a clinical condition which can occur in any workplace or home environment where there is stress, it is recognized by the World Health Organization (WHO) as a syndrome \cite{who19}. In short term, cognitive fatigue may cause sleeping disturbances, anxiety, irritability and hormonal disturbances and in long run, this may result more severe impacts on health safety such as cardiovascular, gastrointestinal and neuropsychological disorders \cite{walker20}.

Current frameworks for cognitive fatigue estimation are mostly self-reported questionnaire based \cite{1, 14}, which is impossible to generate continuous fatigue report by avoiding recall bias \cite{1}. Recent advancement of wearable physical and physiological sensor technologies enable accurate estimation of cognitive fatigue related partial outcomes such as stress, anxiety, sleep quality, mobility etc, which provides ultimate opportunity to researchers to estimate cognitive fatigue continuously \cite{5, 7, 8, 12, 17, 20, 25} that includes actigraphy \cite{21, 34}, heart rate (HR) \cite{4, 21}, Electrocardiography (ECG) \cite{11}, Electroencephalography (EEG) \cite{23} and Electromyography (EMG) \cite{15} sensors along with traditional and deep machine learning techniques. Combining accelerometer with ECG has been a successful attempt as well before \cite{bai20} which proposed to use deep learning frameworks (LSTM with Consistency Self-Attention, LSTM-CSA) but suffers with the lack of adaptability across diverse population.

Due to the dissimilarities among different individual group's responses on cognitive fatigue in terms of physical and physiological contexts, current wearable cognitive fatigue estimation research is constrained in group specific cognitive fatigue estimation [cite]. For example, fatigue detection of video game players \cite{gamers_fatigue}, athletes \cite{gamers_fatigue}, basketball players or heavy exercise performers \cite{exercise_fatigue}. As per many clinical psychologists and mental health researchers, cognitive fatigue estimation should be more personalized, rather than generalized on specific group of people to keep mental healthcare systems sustainable for future generations \cite{generalized_fatigue}.

While, the emergence needs of building personalized cognitive fatigue estimation tool, we have the following key question: \emph{can we develop personalized cognitive fatigue assessment tool considering activity as their activity as domain invariant feature and fatigue as personalized response on each activity?} As we know, autonomic nervous system (ANS) restrains the body's major physiological activities including the heart rate (HR) and gland secretion or electrodermal activity (EDA) \cite{chase_16}. However, these responses are contaminated with physical activity artifacts significantly \cite{mobiquitous20}. The central {\bf hypothesis} of the this paper is: \emph{each performed activity context generates similar artifacts on same activity over diverse population, thus, we can align similar activities (activity-awareness) as person invariant feature and its physiological responses as personalized fatigue feature}. For example, in Fig. \ref{fig:example}, we illustrate the EDA responses on two different activities: (i) steady hand (ii) waving hand over two different physiological states: (i) stress and (ii) no stress. The Fig. \ref{fig:example} clearly shows that same activity (waving hands) has similar EDA response patterns (but different amplitude) due to similar artifacts which signifies our hypothesis.

In this paper, we develop a novel Activity-Aware Recurrent Neural Network (\emph{AcRoNN}) model and utilize it to design a personalized cognitive fatigue assessment framework, that provides the following {\bf key contributions}

\begin{itemize}
\item We develop a novel Activity-Aware Recurrent Neural Network (\emph{AcRoNN}) framework that is able to exploit contextual cues present in any event from actigraphy sensor and then assess cognitive fatigue from physiological (EDA and HR) sensor signal using a deep recurrent neural network.
\item Apply \emph{AcRoNN} on two publicly available data and evaluate the capability of \emph{AcRoNN} framework to improve cognitive fatigue assessment.
\end{itemize}

\begin{figure}[ht!]
 \begin{center}
   \includegraphics[width=\linewidth]{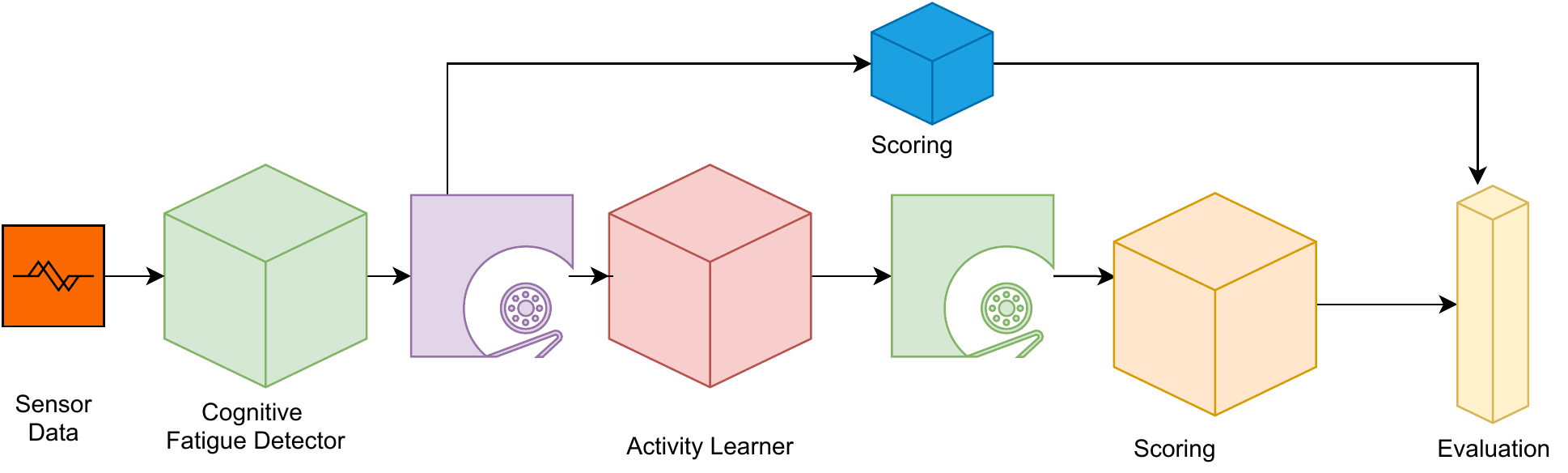}
   \caption{Proposed Activity-Scoring and LSTM-CSA (Consistency Self-Attention) based two stage \emph{AcRoNN} model}
   \label{fig:AcRoNN}
 \end{center}

\end{figure}

\section{Activity-Aware Recurrent Neural Network (\emph{AcRoNN})}
Fig. \ref{fig:AcRoNN} shows the overall schematic diagram of our context-aware cognitive fatigue assessment framework. In this framework, recognized activity information is intertwined with the cognitive fatigue assessment architecture. In this regard, we take (stage 1) cognitive fatigue assessment scoring in terms of gestural activity, and then postural activity information, and finally we (stage 2) re-evaluate the cognitive fatigue assessment scoring based on activity relationships it has learned from stage 1.
\subsection{Stage One: Cognitive Fatigue Detection and Contextual Feature Map Generation}
This stage involves feature extraction, activity recognition and activity-based score mapping for cognitive fatigue detection.
\subsubsection{Wearable Sensor Signal Processing}
Wearable sensors can be two types: physical and physiological. Physical sensors (accelerometer, gyroscope etc.) signal values change over the movements of the sensor devices. Physiological sensors change over physiological condition of body such as EDA changes over stress and PPG changes over heart rate. However, physical movements also impose noises on physiological sensor signals which is called \emph{motion artifacts}.

{\bf Physiological Signal Processing}: A continuous and descrete decomposition of EDA, and time and frequency domain analytics of EDA signal have been investigated before to extract relevant physiological features which were contaminated with noises and motion artifacts \cite{alam16}. \cite{setz10} denoised and classified EDA from cognitive load and stress with accuracy higher than 80\%. Though motion artifacts removal techniques such as exponential smoothing and low-pass filters provide significant improvement in filtering EDA signals, wavelet transforms offer more sophisticated refinement for any kind of physiological sensors such as electroencephalogram, electrocardiogram \cite{greco}, and PPG \cite{lee03}. \cite{chen15} proposed a stationary wavelet transform (SWT) based motion artifacts removal technique. `cvxEDA' proposed  a convex optimization technique considering EDA as a mixture of white gaussian noise, tonic and phasic components where white gaussian noise includes motion artifacts and external noises \cite{greco}. We combine SWT and `cvxEDA' together to remove noises and motion artifacts from EDA signal. Researchers proposed different methods such as frequency analytics \cite{wang13}, statistical analytics \cite{peng14} and digital filter \cite{lee10} to reduce noises and motion artifacts from PPG. We used Periodic Moving Average Filter (PMAF) in this regard \cite{lee07}. After the noise reduction, we generated 33 heart rate variability (HRV) features from PPG (as per \cite{bai20}) and 12 statistical features from EDA (as per \cite{chase_16}) signals with a 10-seconds window.

{\bf Accelerometer Signal Processing}: We used Bai et. al. proposed accelerometer signal processing method \cite{bai20}. We used ActiLife tool \cite{actilife}, and calculated the Actigraphy counts (from accelerometer) every 5 seconds, and detect the non-wear time (for invalid data removal). Within every 10-seconds window, based on Actigraphy counts we further extracted 8 statistical features, i.e.,mean, median, standard deviation, variance, minimum value, maximum value, skewness and kurtosis for further processing.

{\bf Multimodal Feature Sequence Construction}: After preprocessing and feature engineering, the original segment can be transformed into a D-dimensional sequence $X=\{x_t \in \mathcal{R}^D\}^T_{t=1}$ where $T$ is the sequence length (i.e., the number of windows/epochs within a segment), and $x_t = {feat_{acc} \cup feat_{eda} \cup feat_{hrv}}$ where $feat_{acc}, feat_{eda}, feat_{hr}$ represent accelerometer, EDA and HR features extracted above. Since, each of the extracted features were in 10-seconds window, the concatenated input feature $x_t$ has a dimension of 53 ($33+12+8$) per 10-seconds in the time-series window.

\subsubsection{Activity Recognition Module}
We develop a two step multi-label activity recognition framework which consists of two LSTM with Consistency Self-Attention (LSTM-CSA) \cite{lstm_sa} models, (1) gestural activity recognition and (2) postural activity recognition. Both of the LSTM-CSA models are independent from each other, trained and tested separately using hand gesture and postural activity labels respectively using the input accelerometer features ($feat_{acc}$) and their corresponding labels. For all LSTM-CSA models, we used the following regularisation term
\begin{equation}
\Gamma(\alpha) = T\sum_t |\alpha_t - \alpha_{t-1}|
\end{equation}
where $T$, $\Gamma(\alpha)$ tends to penalize heavily with a larger contextual scores (which will be fined later) to maintain its global consistency.
\subsubsection{Class Contextual Feature Maps}
We develop a contextual feature mapping for each cognitive fatigue label which can be represented as follows
\begin{eqnarray}
cfm_c(x,y) = \sum_{cls(b)=c} hm_b(x,y)\nonumber\\
CFM_c = \frac{cfm_c}{max(cfm_c}
\end{eqnarray}
The output of the contextual feature map layer is a ($H_c \times W_c \times C$) tensor, where $H_c$ and $W_c$ are the segment dimensions, and $C$ is the number of classes. We consider (after trial and error), we have set $H_c=23$ while we have already defined $W_c=53$.
\subsubsection{Contextual Scoring}
We designed a scoring function to measure the contextual relevance of cognitive fatigue detection in relation to multiple-label's presence in a window \cite{kevin19}. Scores are computed using the contextual feature maps generated in each stage of our pipeline, and are used as the ranking score in AP calculations to measure whether contextual learner confirms or refutes detections passed to it, based on learned semantic relationships. The scoring process is designed as a new network layer, and appended to the end of each stage in our pipeline. We defined two contextual-scoring method as per \cite{kevin19}.
\begin{itemize}
\item {\bf Individual Contextual Scoring}: Gestural and postural-based cognitive fatigue scoring has been estimated using the following equation
\begin{equation}
Score_1(b) = \frac{\sum FM_b(x,y)}{2\sigma^2_{bx} \times 2\sigma^2_{by}}
\end{equation}
where $FM_b$ represents the activity bounding box (start and end of an activity) related relevance score and $b$ represents each activity type i.e. gestural or postural activity as per \cite{kevin19}. We have two types of individual contextual scoring in our framework, gestural-based cognitive fatigue scoring and postural-based cognitive fatigue scoring (Fig. \ref{fig:AcRoNN}).
\item {\bf Cumulative Contextual Scoring}: In this scoring method, we add both gestural and postural activity based cognitive fatigue scoring together for producing final Activity-based cognitive fatigue re-scoring which can be defined as follows
\begin{equation}
Score_2(b) = \frac{\sum CFM_b(x,y)}{2\sigma^2_{bx} \times 2\sigma^2_{by}}
\end{equation}
where $CFM_b$ represents the cumulative activity bounding box (start and end of an activity) related relevance score and $c$ represents cumulative activity type i.e., either gestural or postural activity-based cognitive fatigue scoring or re-scoring \cite{kevin19}. 
\end{itemize}

\subsection{Stage Two: Activity-Aware Cognitive Fatigue Learner}
The second stage is an LSTM with Consistency Self-Attention (LSTM-CSA) model that is trained to learn semantic relationships using the cumulative contextual score mapping generated by the primary cognitive fatigue detector using the Equation 1.
\section{Experimental Evaluation}
In this section, we aim to evaluate our proposed Activity-Aware Recurrent (\emph{AcRoNN}) performance towards developing a personalized cognitive fatigue assessment system using wearables without any target labels.
\subsection{Datasets}
We use two datasets to evaluate \emph{AcRoNN} model performance which are described as follows:
\begin{itemize}
\item {\bf A1: Activity Recognition Dataset}: Previously, we collected hand gestural (8-hand gestures) and postural (4 postural) activity dataset to serve the purpose of our previous papers \cite{fairness,mobiquitous20} using Empatica E4 watch. We utilized the same dataset and developed hand gesture and postural activity recognition framework as per our proposed framework in this paper.
\item {\bf D1: Gamer's Fatigue Dataset}: We recruited 5 student video games players (age ranges from 19-25) for 7 days who stayed up during a 22 hour shift every alternative day (4 days each) to simulate cognitive fatigue while wearing Empatica E4 watch \cite{empatica}. Empatica E4 watch consists of accelerometer (ACC), electrodermal activity (EDA), photoplethysmography (PPG) and skin temperature (TEMP) sensors. During the data collection (including non-gaming days), participates were asked to measure their sleepiness based on the `Stanford Sleepiness Scale' (SSS) \cite{sss,sss1}  (ranges 1-7 representing active to extremely sleepy) and the `Sleep-2-peak' score \cite{sleep2peak} (ranges 1-7 representing active to extremely sleepy) using Sleep2Peak Android App \cite{sleep2peakapp,s2papp}.
\item {\bf D2: Healthy Adults Fatigue Dataset}: We have used publicly available health adults fatigue dataset \cite{luca20,luo20}. Data from 28 healthy individuals (26–55 years of age, average age 42 years, 41/51\% female/male), of which 17 enrolled up to 2 days after returning from long-haul flights with 3–7 time zone differences and hence were recovering from jet lag, from 1 to 219 consecutive days (mean 35, median 9, total 973 days) were collected. Objective data were collected using a multisensor wearable device, Everion (Biovotion AG, Switzerland \cite{biovotion}), in conjunction with a mobile app, SymTrack (Gastric GmbH, Switzerland), to deliver a daily fatigue questionnaire. Volunteers were asked to continuously wear the Everion device around their non-dominant arm over a 1-week period. The device combines a 3-axis accelerometer, barometer, galvanic skin response electrode, and temperature and photo sensors. Dataset tracked a total of 12 parameters at 1-Hz temporal resolution on physical activity and physiology. Volunteers were instructed to complete a 4-item daily questionnaire in the evening to capture their subjective assessment of fatigue, adapted from the Fatigue Assessment Scale \cite{mich03} and Visual Analogue Scale to evaluate fatigue severity \cite{lee91}: (10 Physical fatigue score (PhF), (2) Mental fatigue score (MF), (3) Visual analogue scale score (VAS), and, (4) Indicator of relative perception (RelP) (see \cite{luca20,luo20} for more details). 
\end{itemize}
\subsection{Pre-Processing}
For each subject and parameter, we excluded days where more than 80\% of the samples were missing to ensure an acceptable performance of downstream analysis. Missing samples were due to subjects not wearing the device (e.g., during charging) or low-quality segments (e.g., loss of skin contact). This filtering step led to a total of 5 subjects and 821 hours of data annotated (Stanford Sleepiness Scale and Sleep-2-peak) labels with continuous Empatica E4 sensor data (excluding 1 hour recharging sessions). Finally, we imputed missing data gaps using the state-of-the-art unidirectional uncorrelated recurrent imputation model from Cao et al \cite{cao18}.
\subsection{Baseline Algorithm development}
We re-implemented latest cognitive fatigue estimation framework \cite{bai20}. In Bai et. al. \cite{bai20}, authors generated highest accuracy of cognitive fatigue assessment using LSTM-CSA model. Although, Bai et. al. provided cognitive fatigue detection method based on PPG (ECG) and Accelerometer sensor signal processing, we implemented the following baselines from Bai et. al. method as follows.
\begin{itemize}
\item {\bf B1:} In this framework, we considered original Bai et. al. approach i.e. using only ECG and Actigraphy features (total number of features 41 = 33+8) and followed Bai et. al. to re-implement original baseline paper's result without any alteration.
\item {\bf B2:} In this framework, we considered our 53 features that include Actigraphy, PPG and EDA features and applied Bai et. al. proposed LSTM-CSA model for cognitive fatigue assessment.
\item {\bf B3:} In this framework, we combined our activity recognizer produced hand gesture and posture detections, and applied Bai et. al. considered 41 features (Actigraphy and ECG) and used our proposed \emph{AcRoNN} framework.
\item {\bf AcRoNN}: We combine everything together as proposed in this paper.
\end{itemize}
\subsection{Results and Comparisons}
Table \ref{tab:result-table} shows details of our experimental results and comparisons with our different baseline models. We can easily identify that our model \emph{AcRoNN} outperforms all of the baseline models significantly in both of our collected datasets and already available datasets. Also, we can firmly say that, our  \emph{AcRoNN} model outperforms baseline significantly even though we chose to use baseline proposed sensors (Actigraphy and ECG) related features only.

\begin{table}[!h]
\begin{scriptsize}
  \caption{Cognitive Fatigue Detection Performance Comparisons across baseline methods}
  \label{tab:result-table}
  \centering
  \begin{tabular}{|p{0.8cm}|p{1.4cm}|p{1.3cm}|p{1.5cm}|p{1.5cm}|p{1.3cm}|p{1.3cm}|p{1.3cm}|p{1.3cm}|}
      \hline
 {\bf Data D1}     & {\bf B1 (D1)}  & {\bf B2 (D1)} & {\bf B3 (D1)}  & {\bf AcRoNN (D1)}
    \\
    \hline
    Precision & $69.65\pm.1$  & $70.45\pm.1$ & $79.58\pm.1$ & $83.87\pm.2$\\    \hline
 \hline
     Recall & $68.64\pm.1$  & $71.76\pm.1$ & $80.4\pm.2$ & $82.45\pm.1$\\    \hline
 \hline
     F1 & $69.23\pm.1$  & $74.56\pm.2$ & $79.45\pm.1$ & $83.45\pm.1$\\    \hline
 \hline
    {\bf Data D2}     & {\bf B1 (D2)}  & {\bf B2 (D2)} & {\bf B3 (D2)}  & {\bf AcRoNN (D2)}\\
    \hline
    Precision & $65.36\pm.4$  & $66.45\pm.1$ & $72.56\pm.1$ & $76.76\pm.2$\\    \hline
 \hline
     Recall & $66.42\pm.1$  & $66.45\pm.1$ & $73.45\pm.1$ & $76.34pm.2$\\    \hline
 \hline
     F1 & $67.35\pm.1$  & $67.34\pm.1$ & $74.47\pm.1$ & $77.47\pm.2$\\    \hline
 \hline

  \end{tabular}
      \end{scriptsize}
\end{table}

\section{Limitations and Future Works}
We have collected data from only 5 student volunteers in a limited setting due to the on-going pandemic related lockdown that blocked us from reaching mass population in the campus and dormitory of University of Massachusetts Lowell. However, our unique Activity-Aware attention model has been evaluated on a publicly available data that provides us ample confidence of the efficacy of our developed framework. We also could not validate the activity recognition accuracy in the collected data due to the unavailability of camera data as per the IRB exemption. However, our collected activity data (A1) has been well-validated by our previous researches which were published in top venues \cite{chase_16,mobiquitous20}, that provides us validity of the activity recognition dataset as well as the related outputs that have been used in our \emph{AcRoNN} framework's stage-1. In future, while the lockdown will be ended, we plan to collect more data engaging more students in the campus and out of campus, more likely in real cognitively stressed and fatigue community such as healthcare workers, construction workers and scuba divers.
\section{Conclusion}
To develop an automated cognitive fatigue assessment system, we introduced a new pipeline from data collection, data preprocessing, feature engineering, attention based LSTM and a novel context-aware LSTM model flow. To our best knowledge, \emph{AcRoNN} is the best cognitive fatigue detection model in the existing literature which can be extended to any other physiological health assessment with proper study design and data collection. Our efficient two-step feature map scoring method provides a new concept in context-aware activity and health monitoring research area that can be utilized to provide appropriate care to patients with dementia, asthma, post-traumatic stress disorder and so on.

\section{Acknowledgment}
We acknowledge Eliza Doering and Alexa Mai for helping us to collect data from students. This project has been funded by University of Massachusetts Lowell Internal Seed Grant to Addressed COVID Related Nursing Community Impact Study. The collected datasets will be made public upon acceptance of the manuscript as per IRB exemption.

\end{document}